\documentclass[a4paper,10pt,twoside]{cpc-hepnp}

\usepackage{multicol}
\usepackage{graphicx}
\usepackage{booktabs}
\usepackage{amssymb,bm,mathrsfs,bbm,amscd}
\usepackage[tbtags]{amsmath}
\usepackage{lastpage}
\usepackage{lineno}
\usepackage{ifpdf}
\usepackage{subfigure}
\ifpdf
\else
\fi

\begin{document}

\fancyhead[c]{\small Chinese Physics C~~~Vol. xx, No. x (2017) xxxxxx}
\fancyfoot[C]{\small 010201-\thepage}


\title{Fast simulation of the CEPC detector with {\textsc{Delphes}}
\thanks{The study is supported by National Key Programe for S\&T Research and Development (Grant NO.: 2016YFA0400400), CAS/SAFEA International Partnership Program for Creative Research Teams, CAS Thousand Talent and Hundred Talent programs, grants from the State Key Laboratory of Nuclear Electronics and Particle Detectors, and National Nature Science Foundation of China (Grant No.: 11675202.)}}

\author{%
Cheng Chen$^{1}$
\quad Xin.Mo$^{2}$
\quad Michele Selvaggi$^{3}$
\quad Qiang Li$^{1}$
\quad Gang Li$^{2)}$\email{li.gang@ihep.ac.cn}%
\quad Manqi Ruan$^{2)}$\email{manqi.ruan@ihep.ac.cn}%
\quad Xinchou Lou$^{2,4)}$%
}
\maketitle

\address{%
$^1$ Institute of High Energy Physics (IHEP), Chinese Academy of Sciences, Beijing 100049, China\\
$^3$ European Organization for Nuclear Research (CERN), Switzerland \\
$^2$ Peking University, Beijing, China\\ 
$^3$ University of Texas at Dallas, Richardson, TX 75080-3021, USA
}

\begin{abstract}
Fast simulation tools are highly appreciated in particle physics phenomenology studies, especially in the exploration of the  physics potential of future experimental facilities. 
The Circular Electron Positron Collider is a proposed Higgs and Z factory that can precisely measure the Higgs boson properties and the electroweak precision observables. 
A fast-simulation toolkit dedicated to the CEPC detector has been developed using {\textsc{Delphes}}.
Comparison shows that this fast simulation tool is highly consistent with the full simulation, on a set of benchmark distributions.  
Therefore, we recommend this fast simulation toolkit for CEPC phenomenological investigations. 
\end{abstract}



\footnotetext[0]{\hspace*{-3mm}\raisebox{0.3ex}{$\scriptstyle\copyright$}2013
Chinese Physical Society and the Institute of High Energy Physics
of the Chinese Academy of Sciences and the Institute
of Modern Physics of the Chinese Academy of Sciences and IOP Publishing Ltd}%

\begin{multicols}{2}

\section{Introduction}\label{sec:intro}

The CEPC\cite{ref:cepc_det, ref:cepc_acc} is a proposed Higgs/Z factory. 
It is expected to collide electron and positron beams at a center-of-mass energy of 91 - 240 GeV.
At 240 GeV center of mass energy, the CEPC will produce 1 million Higgs boson in 10 years.
Operating at 91.2 GeV center of mass energy, the CEPC could produce 10 billion Z boson per year.
Thanks to the large signal statistic and the clean collision environment, the CEPC could boost the precision of Higgs properties measurements by roughly 1 order of magnitude comparing to the HL-LHC, 
and also improve nowadays' accuracy on the electroweak (EW) precision measurement by 1 order of magnitude\cite{ref:cepc_det}. 

Having reliable detector simulation tools is fundamental for the CEPC collaboration. 
Given the large signal statistic at the CEPC (especially during the Z pole operation), 
a fast simulation tool is mandatory for the physics potential exploration. 
The {\textsc{Delphes}~}\cite{ref:delphes} framework has been selected for that purpose by the CEPC collaboration. 
A dedicated CEPC detector model has been designed within {\textsc{Delphes}}, compared and validated against the CEPC Geant4~\cite{ref:geant4} based full simulation. 
In addition, a specific CEPC heavy flavor tagging module has been designed and included in the official {\textsc{Delphes}} release. 
We believe this fast simulation tool could be widely used for CEPC phenomenological studies. 

This paper is organized as following.  
In Sect.{~\ref{sec:detector}}, we introduce the detector used in CEPC full simulation studies, 
and the particle flow principle that governs this current design. 
As a reference, the full reconstruction chain is also presented in detail. 
We also recall the key ideas of the {\textsc{Delphes}} framework and present the CEPC configuration card. 
In Sect.{~\ref{sec:validation}}, we demonstrate the fast simulation performance on a set of Higgs benchmark processes, and compare the key distributions with full simulation.
In Sect.{~\ref{b-tagging}}, the improvement of flavor tagging according to full simulation and its usage are introduced.
The conclusion and discussion are presented in the last section. 

\section{The CEPC detector concept\label{sec:detector} and its core performances}

\subsection{Geometry}

The design of the CEPC detector follows the principle of particle flow. 
The key idea is to follow every final state particle and reconstruct them in the most suited sub-detector system. 
i.e., reconstruct the charged particles in the tracker, veto the calorimeter hits/clusters induced by charged particles and reconstruct the remaining calorimeter signals into neutral particles. 
Therefore, particle flow globally interprets the detector information into reconstructed final particles, which serves as the basis for the physics objects reconstruction. 

Since particle flow algorithm uses coherently all the detector hit information, it provides high efficiency and high purity reconstructed physics objects. 
For taus and the jets, it significantly improves the accuracy in their energy/momentum measurements, 
since the majority of energy of these composed objects is carried by charged particles, 
whose momentum resolution, measured by the tracker, is usually superior comparing to the energy resolution measured by the calorimeter. 

The particle flow oriented detector design exploits the low material budget and high precision tracking system and the high granularity calorimeter. 
The former is required for a precise determination of track momentum, and to limit the chance of nuclear interaction with the tracker material. 
The high-granularity calorimeter plays a key role, to separate and to identify the nature of each cluster. 
Meanwhile, it provides the energy reconstruction to the neutral particles. 
Following these requirements, the CEPC detector design takes the ILD~\cite{ref:ild}, the benchmark detector for the international linear collider~\cite{ref:ilc} studies, as its starting point.
To be adapted to the CEPC collision environment, mandatory changes have been made at this reference detector. 

The CEPC reference detector consists of a hybrid tracking system composed of silicon devices and a Time Projection Chamber (TPC), 
a high granularity calorimeter system, and a superconducting solenoid of 3.5 T and its Yoke system. 
The hybrid tracking system has five parts. A vertex detector (VXD), constructed with high spatial resolution pixel sensor,
is placed very close to the interaction point (IP) and the radius of the innermost layer is 16 mm.
The VXD provides a typically resolution of 5~$\mu$m on the track impact parameters measurements, 
which is used for the $b-/c-$jet flavor tagging and $\tau$-tagging.
A silicon inner tracker cooperates with the VXD for vertex reconstruction and flavor tagging.
A set of forward tracking disks are placed in the forward region
in order to increase the geometric acceptance of tracking system with coverage up to $|\cos\theta| = 0.995$.
A silicon external tracker and end-cap tracking disks are taken as the outermost layer of whole tracker system,
and provide high precision position measurements of tracks entering the calorimetry system.
The TPC, with a 2.35m half-length and 1.8m outer radius, provides over 200 hits per track and 100 $\mu$m resolution in $r\phi$ plane,
which allows for excellent pattern recognition, track reconstruction efficiency, and potential $dE/dx$-based particle identification.

Sampling Electromagnetic Calorimeter (ECAL) and Hadron Calorimeter (HCAL) with very fine granularity are deployed in this reference detector.
Following the ILD design, the ECAL is segmented into 30 layers along the longitudinal direction, each layer consisting of a silicon sensor layer and a tungsten absorber layer. 
The silicon sensor, is then divided into 5~mm by 5~mm readout cells. 
Similarly, the HCAL uses iron absorber and RPC sensors, dividing into 40 longitudinal layers and 10~mm by 10~mm transverse cells. 
A superconducting solenoid of 3.5 T surrounds the calorimetry system.
The flux return yoke and the support structure for the whole detector is placed outside the solenoid. 
Equipped with sensor layers, the yoke system also acts as a Muon system, which significantly increases the sensitive volume of the detector. 

\subsection{Performance}

Being a particle flow oriented detector, the CEPC detector can be characterized by the identification and resolution performance of each individual final state particles and the separation performance. 
The core reconstruction algorithm, Arbor\cite{ref:arbor}, has been optimized for the particle flow oriented detector design at the CEPC. 

Final state particles includes charged particles, photons, and neutral hadrons. 
For the charged particles, the CEPC reference detector provides an efficiency close to 100\% for tracks with transverse momentum larger than 0.1~GeV. 
The typical momentum resolution reaches $\delta(1/P_{T}) = 2\times10^{-5}~\mbox{GeV}^{-1}$ in the barrel region. 
For photons with energy larger than 1 GeV, the CEPC reference detector achieves a relative energy resolution of $20\%/\sqrt{E/\mbox{GeV}} \oplus 1\%$.
For neutral hadrons, the CEPC reference detector delivers an energy resolution of roughly $60\%/\sqrt{E/\mbox{GeV}} \oplus 1\%$.  
 
The separation performance, i.e., the ability of reconstruct nearby particle showers, is a key feature of the Particle Flow Algorithm (PFA). 
This performance can be characterized, at detector level, by the critical distance at which the two nearby showers can be disentangled at clustering. 
Benchmarked with two photon samples, Arbor achieves a critical distance of 9~mm with 5~mm ECAL readout cell size. 
Considering an inclusive $ZH$ sample with the CEPC reference detector, 9~mm critical distance means that 2\% of the final state particles showers (with energy larger than 0.5~GeV) might be overlapped, leading to confusions that degrades the PFA reconstruction performance. 
In other word, at first order, the confusions induced by the shower overlapping at the CEPC reference detector can be ignored. 

Charged particles includes leptons, one of the key physics objects for the CEPC physics program. 
A dedicated lepton identification algorithm, LICH~\cite{ref:LICH}, has been developed. 
For isolated charged particles with energy larger than 2~GeV, LICH reaches a lepton identification efficiency higher than 99.5\% with a misidentification rate of hadron to lepton to be smaller than 1\%.   
The overall Particle Flow performance can also be characterized by the mass resolution of the Higgs decay products in hadronic decay modes. 
For instance, in the $H\to gluons$ channel, a mass resolution better than 4\% is achieved. 

\subsection{Modeling of the performance at {\textsc{Delphes}}}

{\textsc{Delphes}~}\cite{ref:delphes} is a fast simulation framework widely used in phenomenological studies. 
This package is also designed following the principle of particle flow reconstruction, i.e., mimicking the reconstructed final state particles via convoluting the detector effects on top of the MC particles. 
The detector effects includes the detector acceptance via the reconstruction and identification efficiencies, and the energy/momentum resolutions. 
Once final state particles are produced, jets are reconstructed using the standard jet clustering algorithms using the {\textsc{Fastjet}} package~\cite{FastJet}.
In the CEPC cases, since the confusion induced by overlapping can be ignored, the {\textsc{Delphes}} modeling takes parameters according to individual particle reconstruction performances. 
Meanwhile, the geometry dependent parameters such as the acceptance and reconstruction efficiencies have been modeled according to the reference CEPC detector. 
In addition, two modifications has been made and will be possibly integrated to future {\textsc{Delphes}} release. 

The track momentum resolution of the CEPC depends on the polar angle. 
Giving the CEPC detector geometry, the track momentum resolution can be parameterized as a smooth function with respect to the polar angle. 
This modeling has been integrated to the {\textsc{Delphes}}, which improves the modeling of track momentum measurement in the end-caps and the forward region. 

The jet-flavor tagging plays an important role in the CEPC physics program. 
The default flavor tagging modules in {\textsc{Delphes}} only allow to parameterize the tagging efficiency and mis-identification rate at a given working point. 
Given that at the CEPC different physics measurements may use very different working points, a new module has been developed for the fast simulation of the jet-flavor tagging.
This module generates, for any given jet, the likelihoods of being compatible with the b and c flavor hypotheses, according to the flavor of the truth parton and a set of templates derived from full-simulation. 
This module allows the user to easily study the physics performance at different working points.

\section{Validation with full simulation\label{sec:validation}}

The full simulation of the CEPC reference detector has been derived from the ilcsoft framework.
Dedicated digitization modules, mimicking the detector response, produce digitized physical hits according to the MC level hit information. 
Tracks are then reconstructed from the tracker hits. 
The core reconstruction algorithm, Arbor~\cite{ref:arbor}, then interprets the reconstructed tracks and calorimeter hits as as reconstructed final state particles. 

The fast simulation and its reconstruction in {\textsc{Delphes}~} have been validated by comparing the physics object performance on kinematic quantities (such as momentum and angular resolution) and reconstruction/identification efficiencies with full-simulation. 
Several benchmarks channels that provide a exhaustive coverage of the physics objects (jets, leptons and photons) that will be reconstructed with the CEPC detector have been selected for this purpose and will be discussed in what follows.

\subsection{$e^+e^-\to \mu^+\mu^-H$}

The $e^+e^- \to \mu^+\mu^-H$ process is a key channel to be studied at a $e^+e^-$ Higgs factory since it provides a model-independent measurement of the Higgs mass and its production rate.
The di-muons come from Z boson decay and the Higgs is reconstructed from the recoiling system, since the initial four-momentum is precisely defined.
The measurement of the recoil mass distribution provides crucial information on the Higgs particle without detecting directly any of the Higgs decay products. 
Fig.~\ref{fig:mumuh} shows comparisons between the {\textsc{Delphes}~} and full-simulation on the di-muon invariant mass (left) and recoil mass (right)
In both classes, excellent agreement is found. 

\begin{center}
\includegraphics[width=0.49\linewidth]{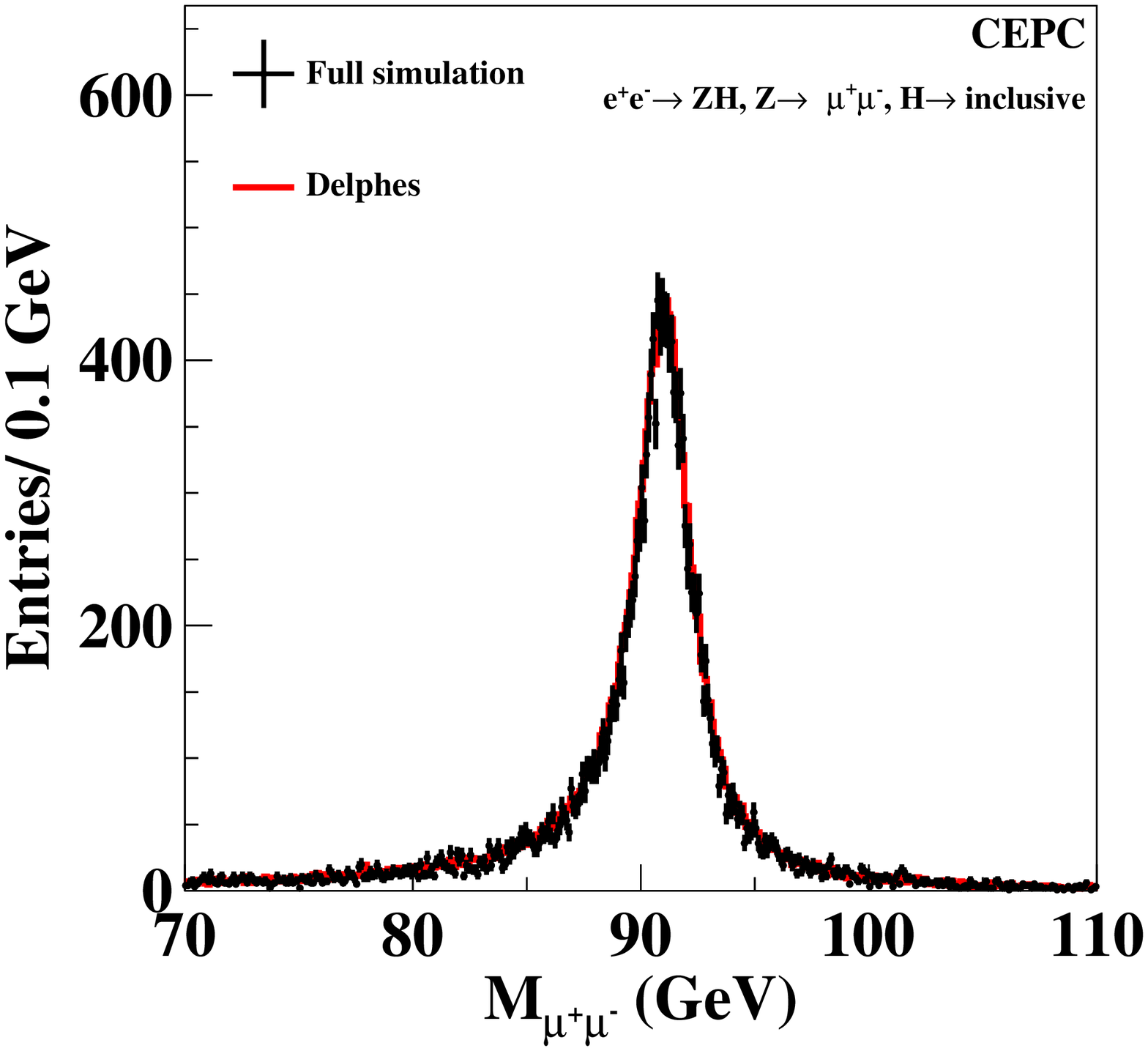}
\includegraphics[width=0.49\linewidth]{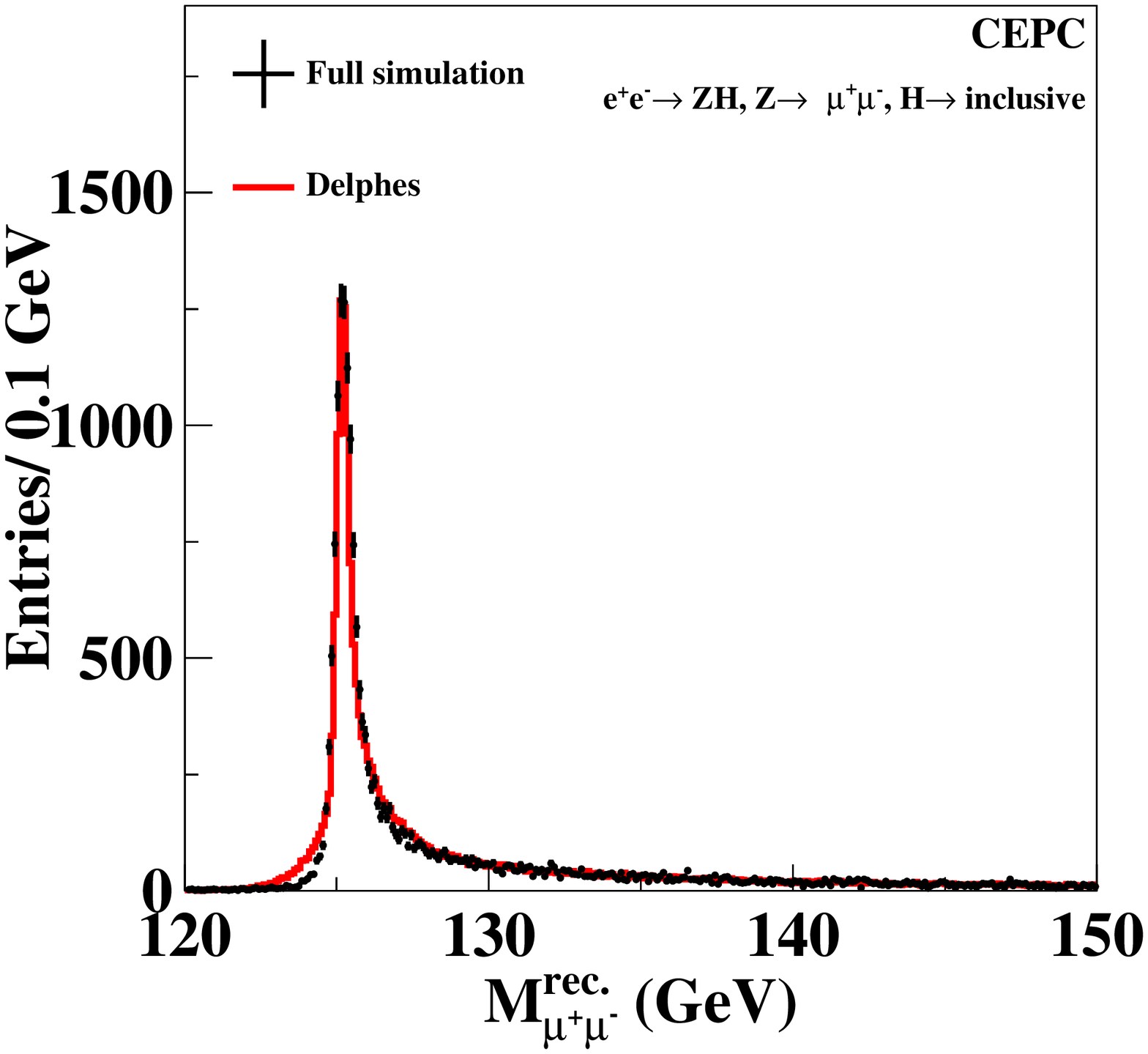}
\figcaption{\label{fig:mumuh}The di-muon invariant mass (left) and recoil mass (right) distribution in $e^+e^- \to \mu^+\mu^- H$ events. The dot with error bar and red histogram represent respectively full and fast simulation (same in the following plots).
}
\end{center}

\subsection{$e^+e^-\to \nu\bar{\nu}H$, $H \to \gamma \gamma$}

The $e^+e^- \to \nu\bar{\nu}H$, $H\to \gamma \gamma$ channel can be used for validation the photon reconstruction with the CEPC detector. 
There are only two isolated photon in the final state if the initial state radiation (ISR) photons are neglected.
Fig.~\ref{fig:hgg} shows the comparison between fast and full simulation,
and it can be seen that the energy resolution of photon is well modeled by fast simulation.
It should be noted that the photon conversion are not taken into account in {\textsc{Delphes}~}.
However, the converted photons can mostly be identified and reconstructed at the CEPC collision environment. 
Thus the degrading due to in-efficient photon reconstruction is rather limited. 

\begin{center}
\includegraphics[width=0.8\linewidth]{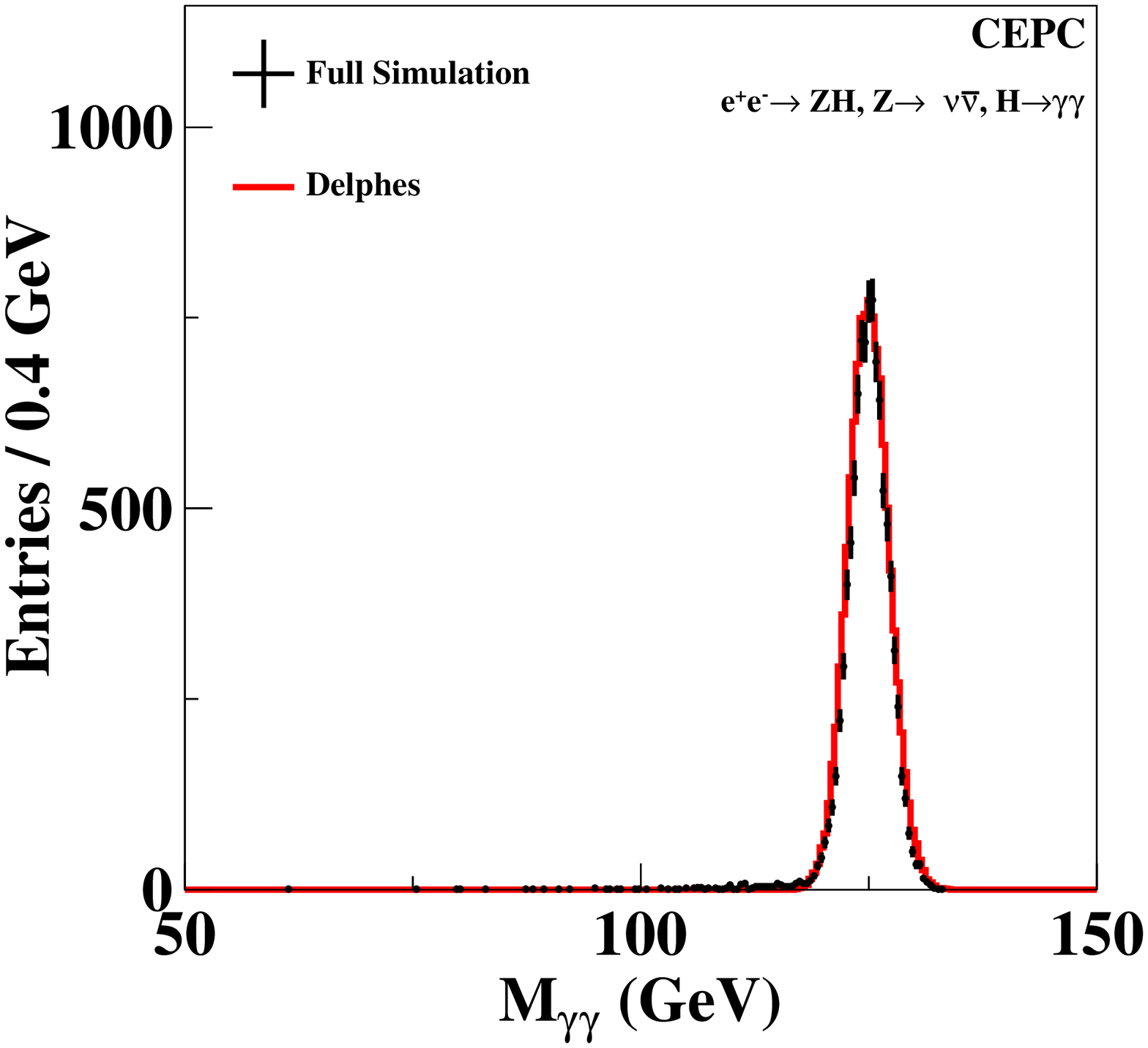}
\figcaption{\label{fig:hgg}The invariant mass distribution of di-photon system. }
\end{center}

\subsection{$e^+e^-\to \nu\bar{\nu}H$, $H \to WW^*$}

The $e^+e^- \to \nu\bar{\nu}H,~H\to WW^*$ process includes the Higgs-strahlung and $WW$ fusion contributions.
This process provides an inclusive test sample since it include the physics objects of jets, leptons and missing energy/momentum in its decay final states. 
Fig.~\ref{fig:nnh} (left) shows the invariant mass of the visible decay products where as Fig.~\ref{fig:nnh} (right) shows the missing mass. 
Excellent agreement is observed between fast and full simulation. 

\begin{center}
\includegraphics[width=0.49\linewidth]{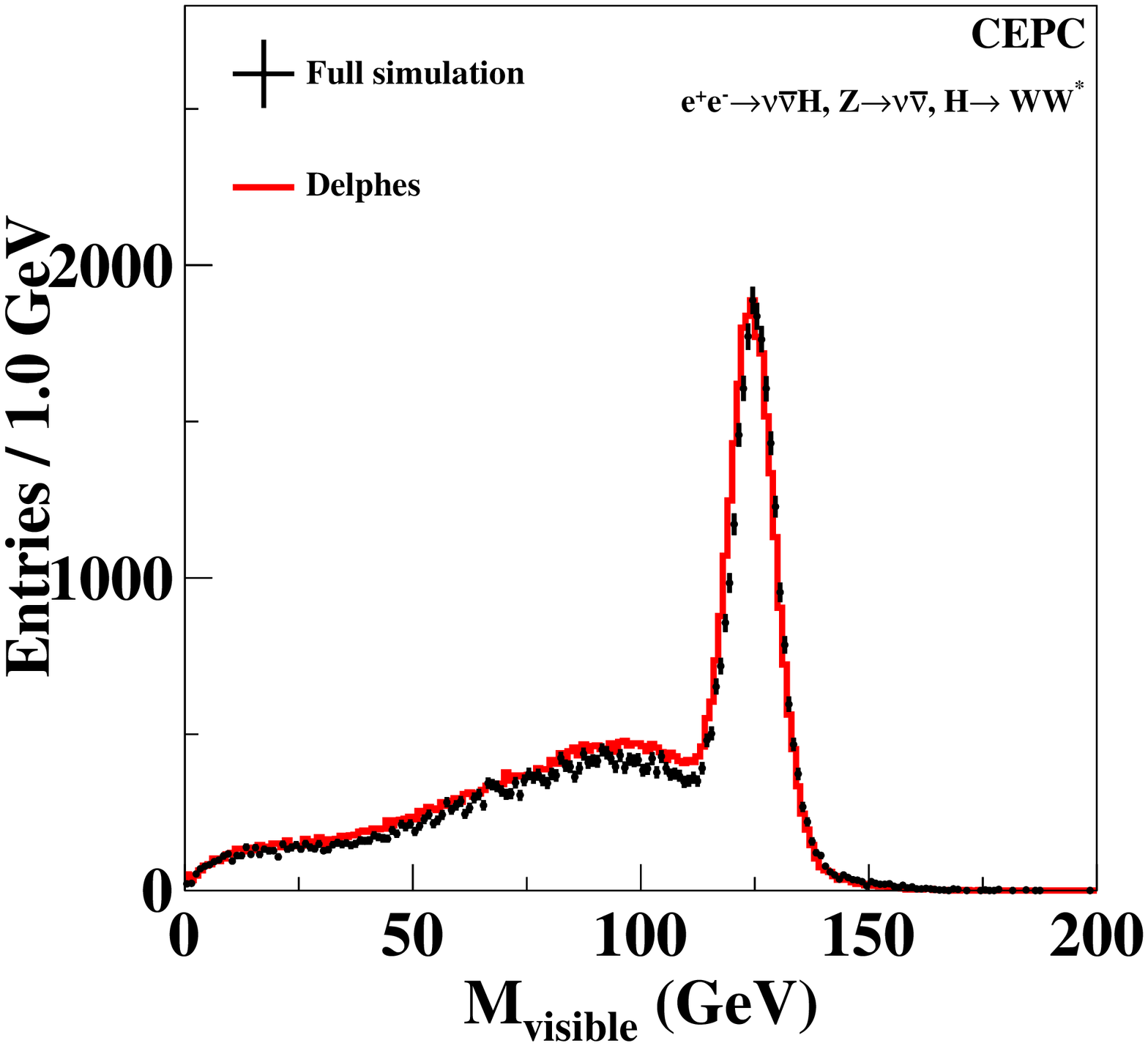}
\includegraphics[width=0.49\linewidth]{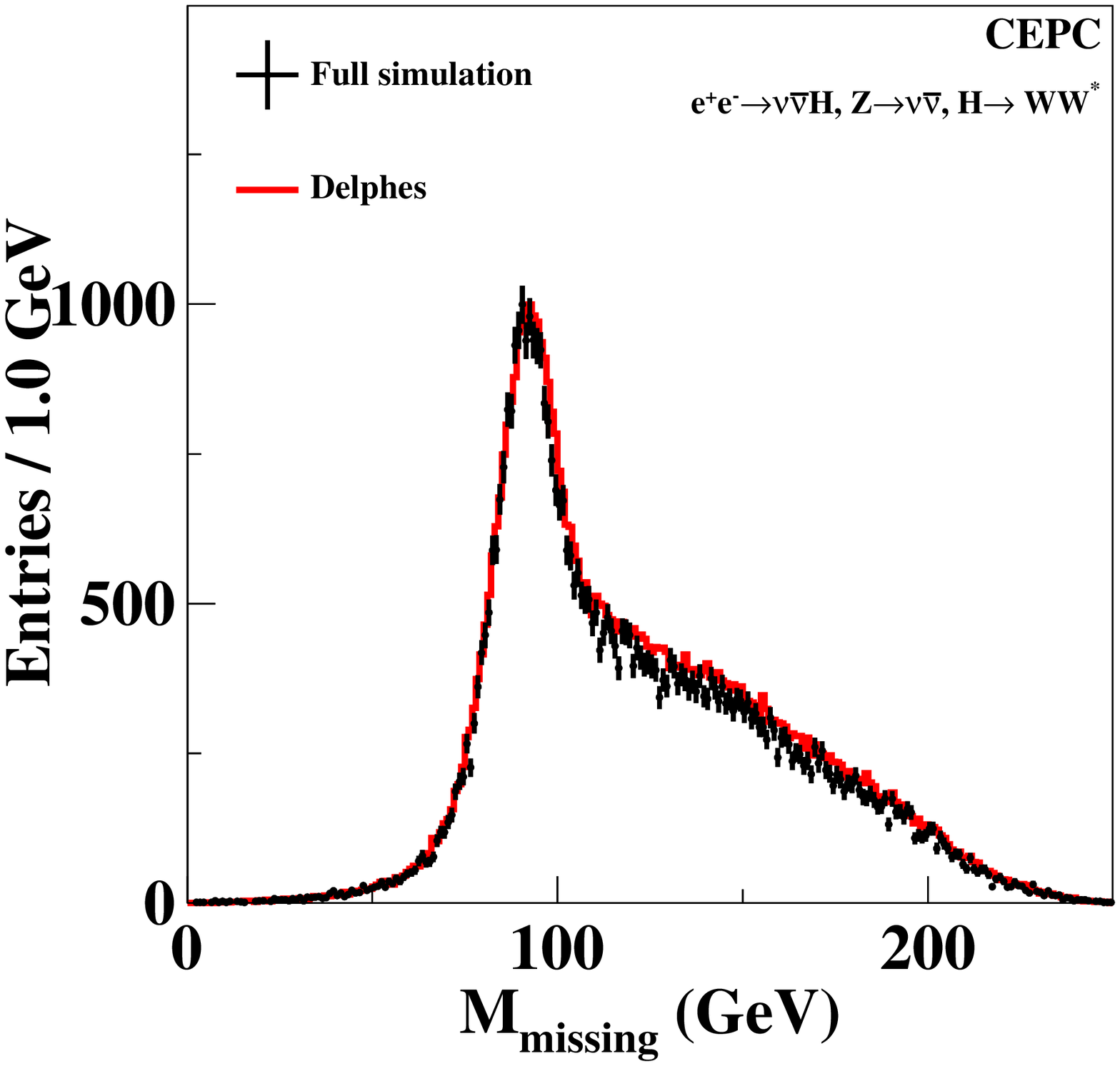}
\figcaption{\label{fig:nnh} 
The visible invariant mass (left) and recoil mass (right) distributions in $e^+e^- \to \nu \nu H, H \to WW^*$ events. }
\end{center}

\subsection{$e^+e^- \to ZH \to 2(q\bar{q})$}

The $e^+e^- \to ZH \to 2(q\bar{q})$ process contains four jets in the final state, and can hence be used for validating jet reconstruction performances, such as jet-clustering, jet energy resolution, and jet pairing.
In both full and fast simulation, all the final states particles are forced into four jets with $ee_{kt}$ algorithm~\cite{ref:eekt}.
The four jets are then paired by minimizing $\chi^2 = (M_{j_1,j_2} - M_Z)^2 + (M_{j_3,j_4}-M_H)^2$,
where $M_Z$ and $M_H$ are fixed to be the world average values~\cite{ref:PDG}. 
The invariant mass distribution of jet pairs are shown in Fig.\ref{fig:zh4q}.
From Fig.~\ref{fig:zh4q}, it can be seen that both of the distributions are highly consistent.
In Fig.~\ref{fig:zh4qsc}, the scattering plots of $M_{j_1,j_2}$  vs. $M_{j_3,j_4}$ 
demonstrate similar patterns in the available kinematic region.

\begin{center}
\includegraphics[width=0.45\linewidth]{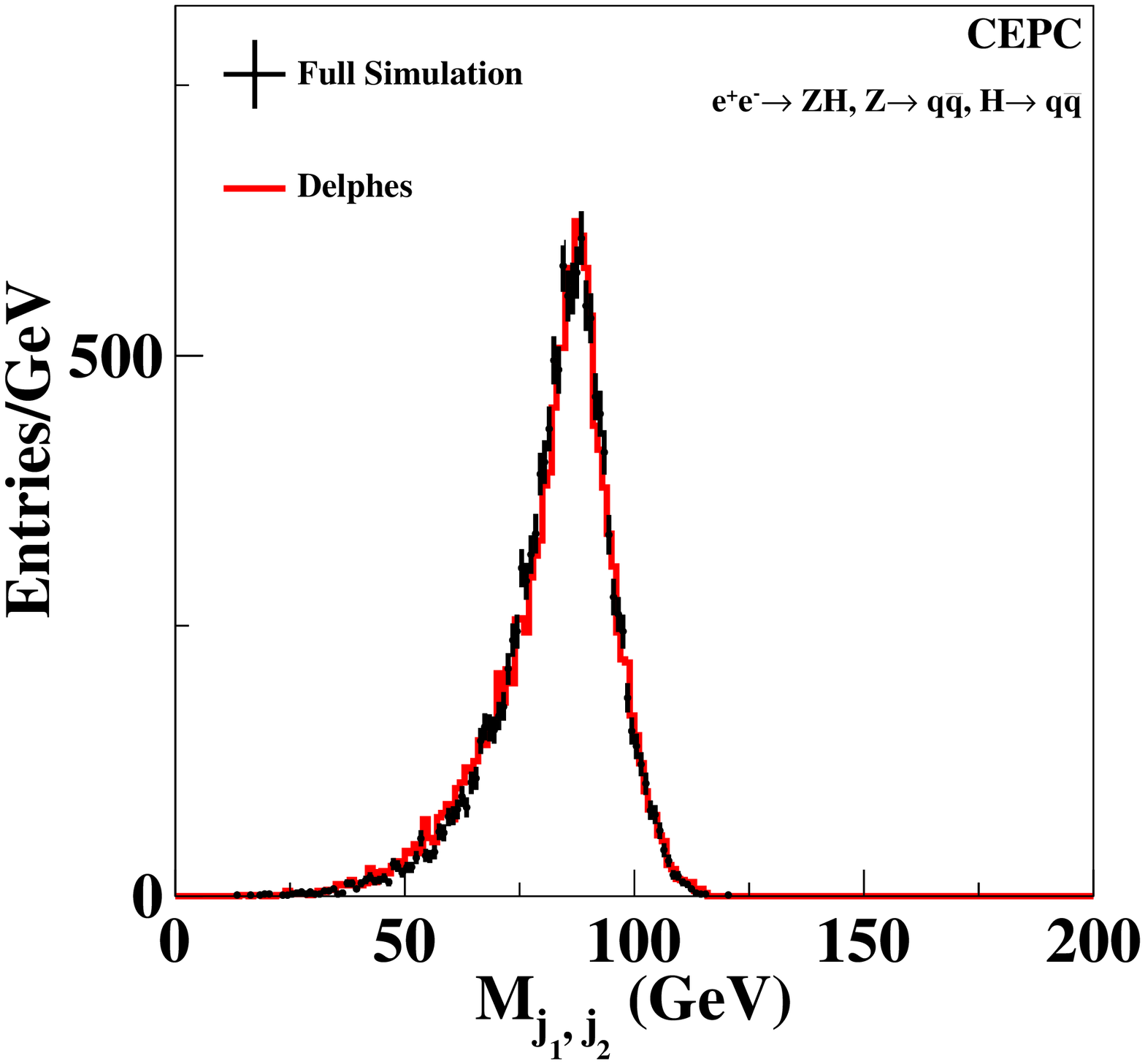}
\includegraphics[width=0.45\linewidth]{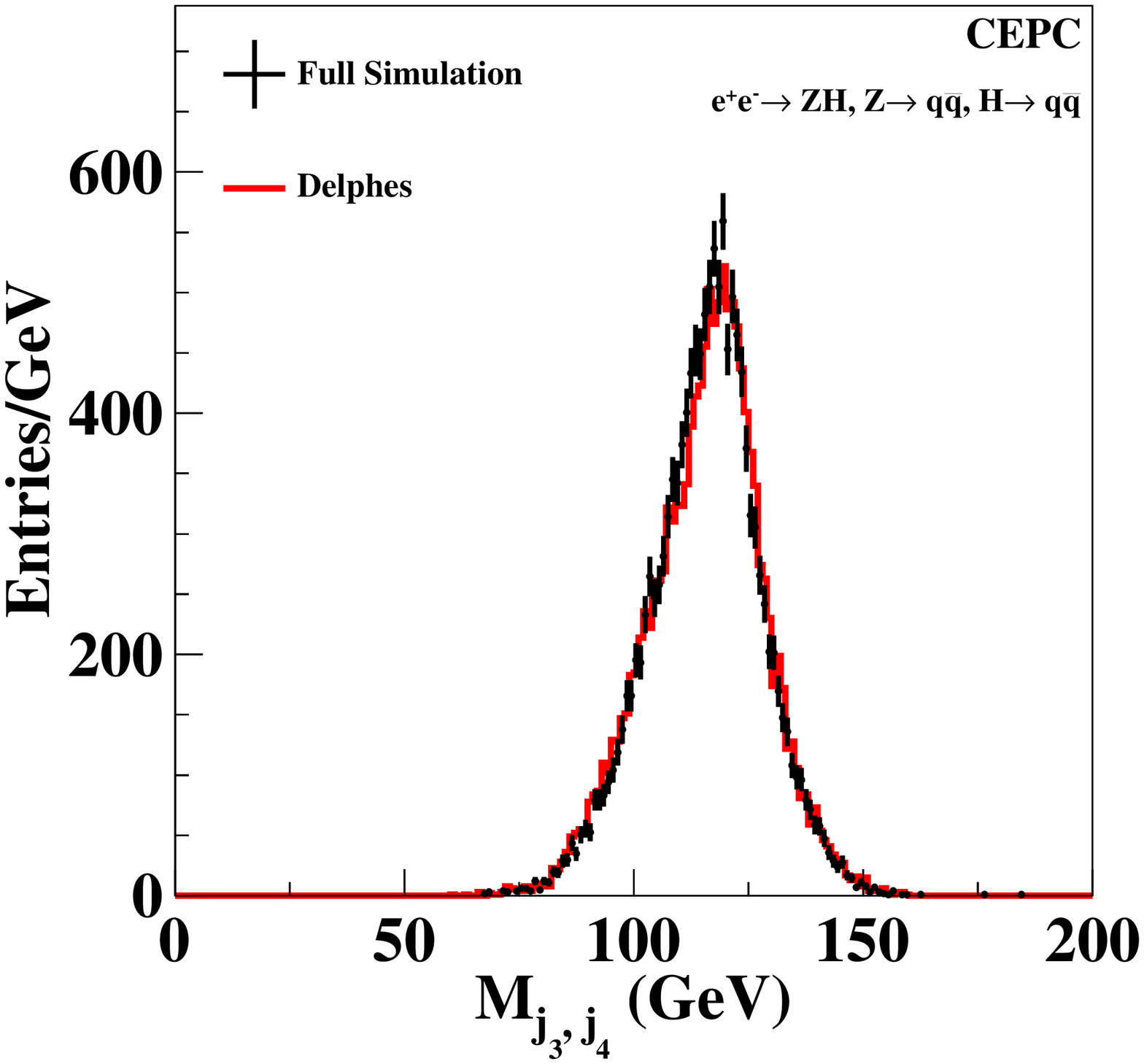}
\figcaption{\label{fig:zh4q} The invariant mass of the jet pair corresponding to the reconstructed Z (left) and Higgs (right) systems.}
\end{center}

\begin{center}
\includegraphics[width=0.49\linewidth]{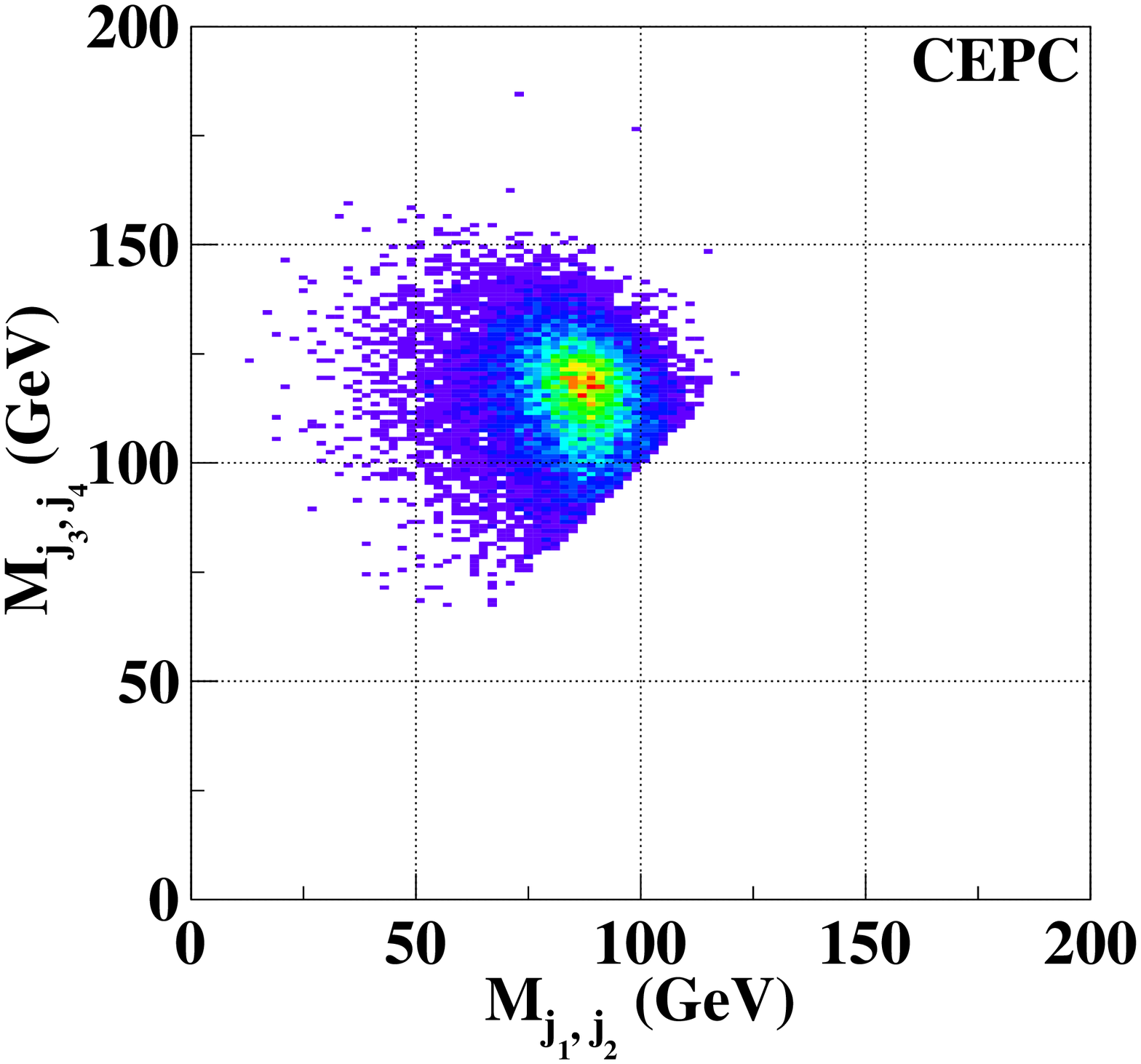}
\includegraphics[width=0.49\linewidth]{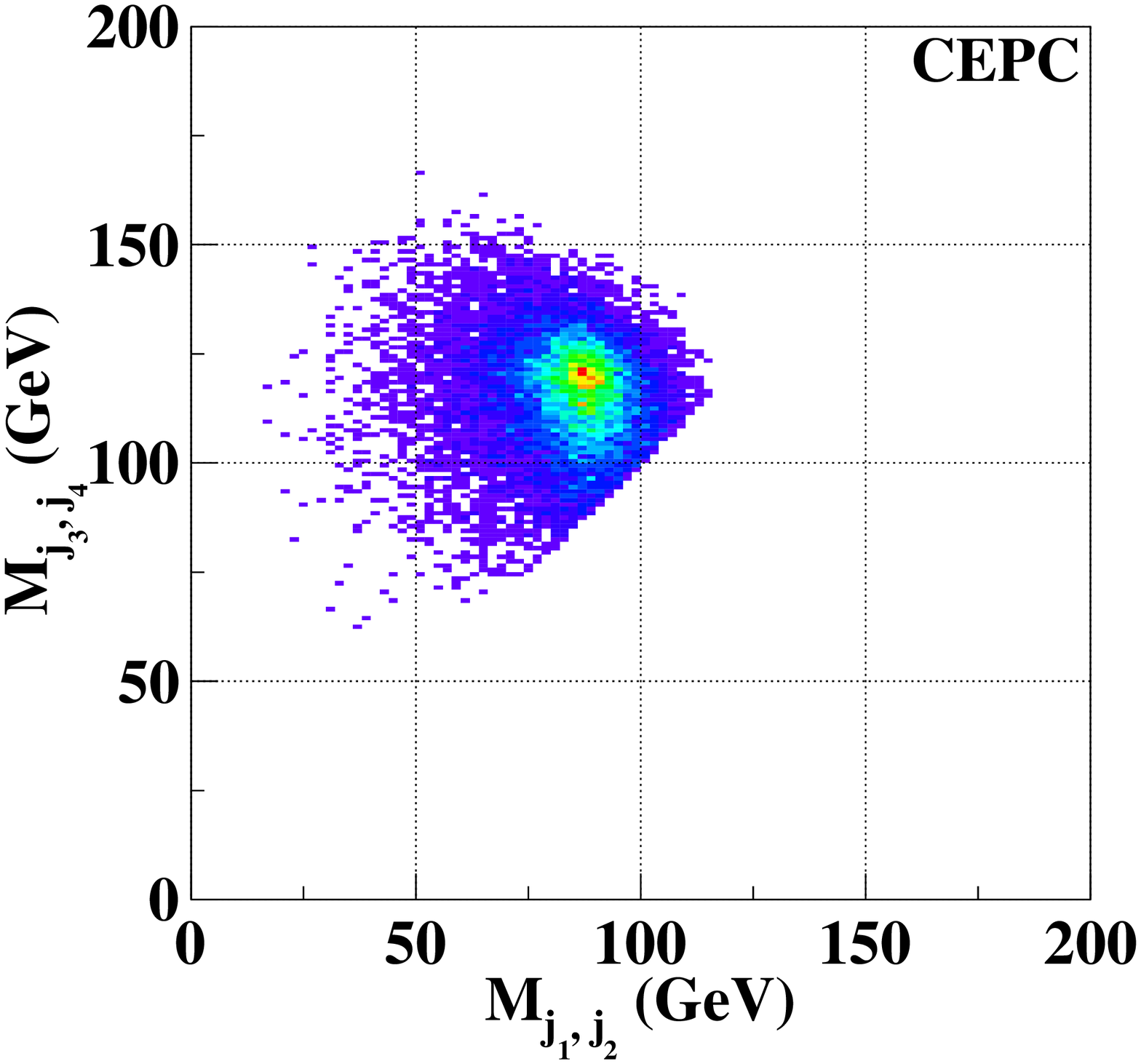}
\figcaption{\label{fig:zh4qsc} The Scattering plots of $M_{j_1,j_2}$ vs. $M_{j_3,j_4}$ in full (the left) and fast (the right) simulation, respectively.}
\end{center}

\section{Flavor tagging based machine learning\label{b-tagging}}

In the original {\textsc{Delphes}} b-tagging module, the jet is tagged as a b-candidate if a b-parton is found within
some distance $\Delta R = \sqrt{(\eta^{\mbox{\scriptsize jet}}-\eta^b)^2+(\phi^{\mbox{\scriptsize jet}}-\phi^b)^2}$ of the jet axis.
The probability to be identified as $b$ depends on user-defined parameterizations of the $b$ efficiency.
The user can also specify a mis-tagging efficiency parameterization.
In the CEPC full simulation, the b and c tagging are based on a multi-class BDT discriminant that is computed via the package LCFIPlus~\cite{ref:lcfiplus},
In this approach, the jet flavor is identified based on a series of observables (up to 60), including secondary vertex mass, jet shape and energy, etc.
LCFIPlus then computes and attaches to each jet the probability for this jet to be a b-jet ($P_b$), a c-jet ($P_c$), and a light jet ($1 - P_b - P_c$).
The output of the above algorithm is included in the {\textsc{Delphes}} output, thanks to a specifically designed module. 

\begin{center}
\includegraphics[width=0.98\linewidth]{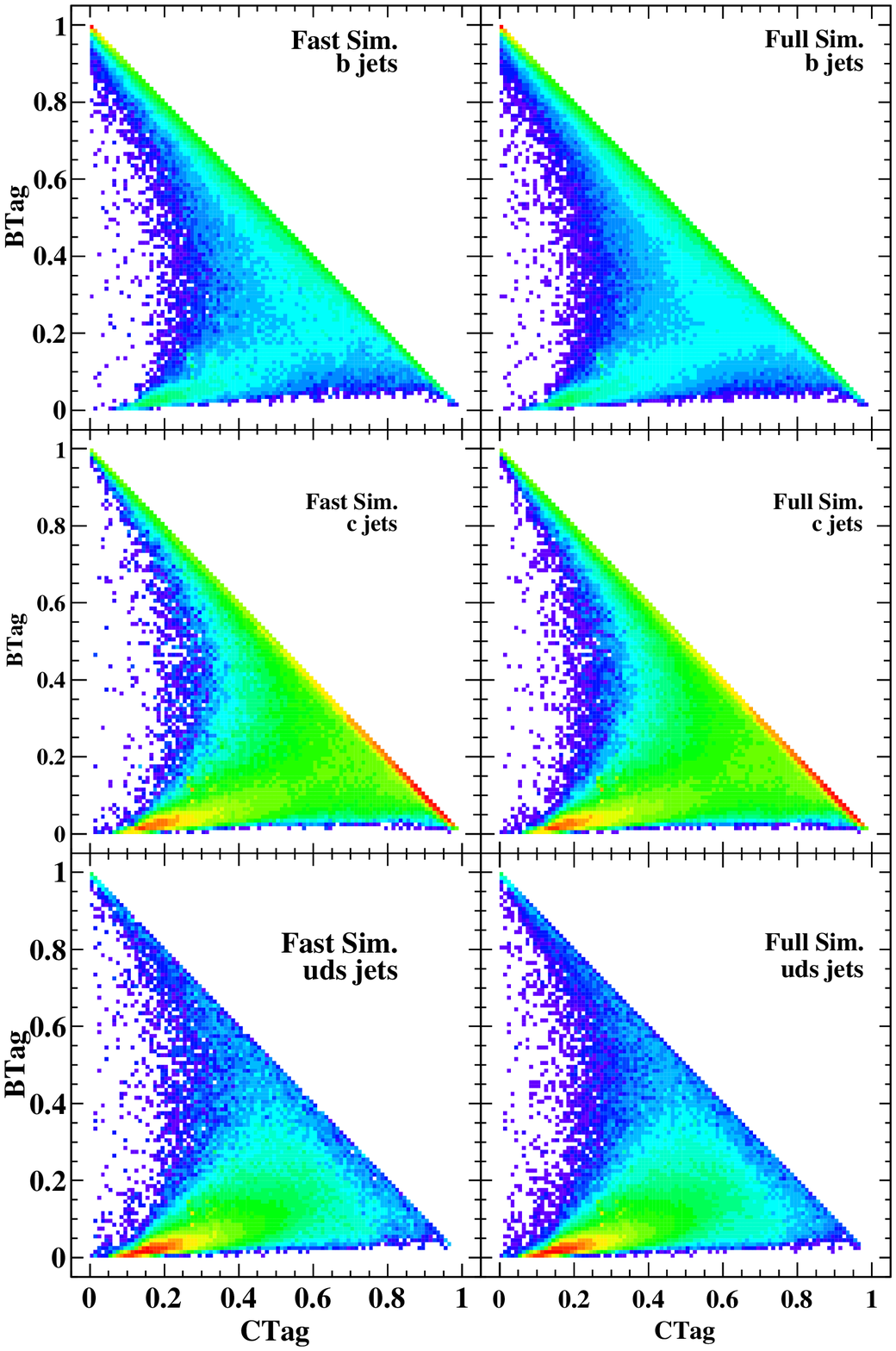}
\figcaption{\label{fig:ft} Two dimensional distributions of $b$-probability versus $c$-probability for $b$, $c$ and light jets, from full simulation(the left) and fast simulation(the right), each with 100k events. }
\end{center}

The up plots in Fig.~\ref{fig:ft} shows the two-dimensional probability $P_b, P_c$ distributions for various jet flavors obtained from full simulation, 
whereas the lower ones are the same distribution as obtained from {\textsc{Delphes}} is shown. 
Again, very good agreement between fast and full simulation is found.

\section{Conclusion\label{sec:conclusion}}

A fast simulation tool to be used for dedicated CEPC phenomenological investigations is needed. 
The flexibility of the {\textsc{Delphes}} framework easily allowed for a configuration of the CEPC detector. 
Comprehensive validation of the performance on key observables have been done. 
Very good agreement has been found between the full Geant4 based simulation and reconstruction and the fast simulation {\textsc{Delphes}}. 



\end{multicols}

\clearpage

\end{document}